\documentclass{article} 

\usepackage{graphicx}

\def\lesssim{\mathrel{\mathpalette\fun <}}
\def\gtrsim{\mathrel{\mathpalette\fun >}}
\def\fun#1#2{\lower3.6pt\vbox{\baselineskip0pt\lineskip.9pt
  \ialign{$\mathsurround=0pt#1\hfil##\hfil$\crcr#2\crcr\sim\crcr}}}

\begin{document}
\begin{flushright}
{\small PM/04-31}\\[.5cm]
\end{flushright}
\begin{center}
{\Large\bf Natural gravitino dark matter in $SO(10)$ gauge mediated
supersymmetry breaking}
\bigskip

Martin Lemoine$^{(a)}$, Gilbert Moultaka$^{(b)}$, Karsten Jedamzik$^{(b)}$,  
 \bigskip

{\it (a)  Institut d'Astrophysique de Paris, \\ 
UMR 7095 CNRS, Universit\'e Pierre \& Marie Curie, \\
98 bis boulevard Arago, F-75014 Paris, France}

{\it (b) Laboratoire de Physique Th\'eorique \& Astroparticules, \\ 
UMR 5207 CNRS, Universit\'e Montpellier II, \\
Place E. Bataillon, F-34095 Montpellier Cedex 5,
France}
\vskip 2cm
\end{center}

\noindent
{\bf Abstract.} It is shown that gravitinos with mass $m_{3/2}\sim
0.1- 1\,$MeV may provide suitable cold dark matter candidates in
scenarios of gauge mediated supersymmetry breaking (GMSB) under
$SO(10)$ grand unification coupled to supergravity, which accommodate
a messenger sector of mass scale $M_X\sim 10^6\,$GeV. This is due to
the combined effects of renormalizable loop-suppressed operators and
generic non-renormalizable ones governing the dilution of a
pre-existing equilibrium gravitino abundance via messenger decay. The
above range of gravitino and messenger masses can be accommodated in
indirect GMSB scenarios.  The gravitino abundance does not depend on
the post-inflationary reheat temperature and it is shown that
leptogenesis can generate successfully the baryon asymmetry.
\vskip 2cm

\section{Introduction}

In particle physics models in which the dynamical breaking of
supersymmetry is transmitted to the visible sector through
renormalizable gauge interactions, the so-called GMSB
models~\cite{GMSB0,GMSB,GR99}, the lightest supersymmetric particle is
the gravitino. As such it can in principle make up the cold dark
matter component of the Universe. Previous studies have demonstrated
that the messenger sector that is responsible for the transmission of
SUSY breaking in GMSB models, may play a key r\^ole in the
determination of the relic abundance of
gravitinos~\cite{BM03,FY02,FY02b,FIY04}. In particular, the lightest
messenger particle with mass $M_X \gtrsim 10^5\,$GeV is assumed to be
stable in GMSB models proposed so far. However messenger number must
be violated in some way, otherwise this lightest messenger would
overclose the Universe \cite{JLM04} (except for the particular case
$M_X\sim10^4\,$GeV \cite{DGP96}, \cite{HMR05}). On general grounds one expects
this messenger to visible sector coupling to be suppressed in order to
preserve the successful phenomenology of GMSB models, in particular
with respect to the natural suppression of flavor changing neutral
currents and correct electroweak breaking~\cite{DNS97}. If the decay
width of the lightest messenger is sufficiently suppressed, the decay
may be a significant source of entropy which reduces the gravitino
abundance.

  Following this line of thought, we show in the present Letter that
the gravitino can indeed provide a natural cold dark matter candidate
in the ``simplest'' GMSB models under $SO(10)$ grand unification
coupled to supergravity. More specifically, we find that for a
gravitino mass $m_{3/2} \sim 1\,$MeV and $M_X\sim 10^6\,$GeV, generic
non-renormalizable messenger-matter couplings allow to obtain the
right relic abundance for dark matter independently of the
post-inflationary reheat temperature. Leptogenesis can thus operate
successfully at high temperatures and produce the observed baryon
asymmetry. Such gravitino and messenger masses are predicted in
indirect GMSB models such as proposed by Dine and
collaborators~\cite{GMSB}. In these models, dynamical breaking occurs
at scale $\Lambda_{\rm DSB}$ in a secluded sector and results in $M_X
\sim \kappa \Lambda_{\rm DSB}$, with $\kappa< 1$ denoting a
combination of model dependent coupling constants, and $m_{3/2}\sim
\Lambda_{\rm DSB}^2/m_{\rm Pl}$. Hence for $\Lambda_{\rm DSB}\sim 10^6
- 10^7\,$GeV, one obtains $M_X\lesssim 10^6\,$GeV and $m_{3/2}\sim
\lesssim 100\,$keV. For the same values, one also finds a
fermion-boson squared mass splittings in the messenger sector $F_X
\sim M_X^2$ which results in correct sfermion, squark and gaugino
masses $m_{1/2} \sim m_0 \sim (\alpha/4\pi)
F_X/M_X \sim 1$TeV~\cite{GMSB,GR99,DTW97}.

\section{Gravitino abundance}

We assume that the post-inflationary Universe reheats to high
temperature $T_{\rm RH}\gtrsim 10^{10}\,$GeV as generically occurs in
scenarios where the inflaton couples through renormalizable
interactions to the visible sector. Then sparticles as well as
messengers are initially brought to thermal equilibrium. Goldstinos,
or equivalently helicity $\pm1/2$ gravitinos, are also brought into
thermal equilibrium as the temperature exceeds the threshold $T_{\rm
eq}\sim 10^5\,{\rm GeV}\,(m_{3/2}/100\,{\rm keV})^2(M_3/10^3\,{\rm
GeV})^{-2}$ ($M_3$ gluino mass) at which production of goldstinos by
sparticle scattering occurs faster than a Hubble
time~\cite{FY02}. Note that gravitino production by messenger
scatterings may further reduce this temperature~\cite{CHKL99}.
Neglecting the abundance of helicity $\pm3/2$ gravitinos, the number
to entropy density ratio of gravitinos after reheating is thus well
approximated by the thermal equilibrium value $Y_{3/2}^{\rm eq}\equiv
n_{3/2}/s\simeq 1.80\cdot 10^{-3} (g_\star/230)^{-1}$.

If messengers sit in a pair of $\mathbf{16}+\mathbf{\overline{16}}$
spinor representations of $SO(10)$, the lightest messenger is a linear
combination of the $\tilde\nu_R$-like [i.e. $SU(3)\times SU(2)\times
U(1)$ singlet boson] components of the $\mathbf{16}$ and of the
$\mathbf{\overline{16}}$ conjugate,  as was first noted in ref.~\cite{DGP96}.
Since the singlet nature of the lightest messenger is a key issue for
our study, it is important to qualify its genericity. Indeed, 
the mass degeneracy in these representations is lifted by various contributions
to the non-singlet states:
tree-level electroweak $D-$term corrections, renormalization group running 
effects in $M_X$ and $F_X$, as well as genuine electroweak loop corrections. 
In the literature~\cite{DGP96, HH97} the focus was mainly on mass splitting
in $SU(2)$ doublets. Furthermore, the leading effects from the running 
of  $M_X$ and $F_X$ can be milder in the GMSB models under consideration
where these parameters are generated dynamically at scales much below 
$M_{\rm GUT}$.  We have thus recomputed the loop corrections to the (squared) 
masses of the $SU(3)\times SU(2)\times U(1)$ charged scalar messengers, 
both in the supersymmetric ($F_X=0$) and susy breaking ($F_X \neq 0$) 
configurations, in a minimal subtraction scheme, and taking the renormalization
scale $\mu \sim M_X$ so as to resum the leading log effects. These corrections
are found to be positive, both in the $F_X=0$ limit and in most of the
parameter space when $F_X \neq 0$, increasing the masses by $\sim 10-20\%$. 
For instance, in the susy and $M_X \gg m_Z$ limit, the correction to the 
$(\tilde\nu_L, \tilde{l}_L)$-like messenger masses is 
$\frac{3}{2\sin 2 \theta_w} \sqrt{\frac{\alpha}{\pi}} \; M_X$ while a somewhat
larger correction obtains for the $\tilde{l}_R$-like and squark-like 
messengers. These corrections should be added to the renormalization group
effects on $M_X, F_X$ when running from the $GUT$ scale down to the messenger 
scale. When applicable, the latter running largely dominates and increases 
further the masses of the $SU(3)\times SU(2)\times U(1)$ charged messengers 
over that of the singlet by $20-60\%$.
In the sequel we denote by $X$ this lightest singlet messenger.

The relic abundance of this lightest messenger is determined as usual
by annihilation freeze-out from thermal equilibrium (provided the
decay width to the visible sector is sufficiently suppressed, as will
be the case here). Since $X$ is a singlet under the standard model
gauge group, it can annihilate either {\sl i)} at the tree level into
visible sector particles and only through the exchange of a GUT mass
$SO(10)$ gauge boson, {\sl ii)} at the tree-level into a pair of
goldstinos, {\sl iii)} at the one-loop level where the heavy as well
as the light particles of the visible (and spurion) renormalizable sector of 
the model contribute in the loops.

The contribution from {\sl i)} is highly suppressed by a factor
$(M_X/M_{\rm GUT})^4$ and can be safely neglected. Annihilation into
goldstinos {\sl ii)} is induced by supergravitational interactions
(see also \cite{JLM04}). It receives contributions from various sectors
including the purely gravitational $s-$channel exchange of gravitons,
$4-$leg contact interactions between $X$ and the gravitino,
$t-$channel exchange of the fermionic partner of $X$, as well as
$s-$channel exchange of the scalar partners of the fermions which make
up the goldstino after supersymmetry breaking. The coupling of the
latter scalars to the gravitinos is fixed by the super-Higgs
mechanism, \cite{BR89, BR88}, while their coupling to $X$ has also a
renormalizable non-hidden contribution through  the coupling of
the spurion to $X$ in the GMSB superpotential. In contrast to the
case of fermions or gauge bosons where the scale of unitarity
violation is reduced from the Planck scale to much lower scales by the
smallness of the gravitino mass \cite{BR89, BR88, G96}, the cross
section for $XX\rightarrow \tilde G\tilde G$ has a mild high energy
behavior, barring non-minimal K\"ahler contributions which are Planck
suppressed. Furthermore, we consider the configurations where the
spurion is heavier than the lightest messenger scalar, which is indeed
the case for parts of the parameter space.\footnote{For instance, in
the simplest model with an $SU(6) \times U(1) \times U(1)_m$ gauge
group in the dynamical breaking sector~\cite{GMSB}, such
configurations are achieved for $\sqrt{3} \le \kappa/\lambda \lesssim
2.2$ and $\lambda_1/\lambda \ll 1$, where $\lambda$, $\kappa$ and
$\lambda_1$ denote respectively the spurion self-coupling, its
coupling to the messenger fields, and its coupling to the $U(1)_m$
charged fields in the superpotential.  Wider ranges are also found for
larger values of $\lambda_1/\lambda (<1)$.  } In the opposite case a
quick annihilation of the lightest messenger to spurion pairs would
occur through tree-level diagrams [accompanied by one-loop induced
spurion decays mainly into gluons] leading typically to a too low $X$
relic abundance. In the limit of very heavy spurion we find for the
$X$ particles annihilating at rest $\langle\sigma_{XX^*\rightarrow
\tilde G\tilde G}v\rangle\,\simeq\, (1/ 8\pi) F_X^2M_X^2/F^4 $, where
$F\equiv\sqrt{3}m_{3/2}m_{\rm Pl}$ is the goldstino decay constant.

The contribution from {\sl iii)} to the annihilation into standard
model particles is fully controlled by the renormalizable 
sector, but can be comparable to the latter in the range of parameters
we consider.  Though a detailed description of these effects is beyond
the scope of the present paper we point out here some qualitative
features in the case of the leading annihilation into two gluons. 
 In this case there is no tree-level annihilation of the type
{\sl i)} and all contributions proceed via one-loop diagrams.
They originate either from the spurion/messenger sector or from the
$SO(10)$ gauge sector. In the latter case, only heavy particles will contribute
to the loops since the lightest messenger is an MSSM singlet. The GUT scale 
contributions will thus decouple leading to negligible effects, unless the
pattern of $SO(10)$ breaking down to the MSSM contains relatively light 
intermediate gauge sectors \cite{CMP84}. We will assume here 
that such patterns do not occur and neglect the effects of this sector
altogether in the sequel [a detailed study of possible effects 
will be given elsewhere].  

The main contributions to {\sl iii)} will come from the spurion/messenger
sector with s-channel exchange of scalar spurion and one-loop
vertices, and box diagrams with virtual exchange of messenger and
spurion fermions. The renormalizability of this sector, together with the fact
that both the spurion and the lightest messenger
are $SU(3) \times SU(2)_L \times
U(1)_Y$ singlets, garantees the cancellation of all infinities as well as
the renormalization scale dependence. 
The resulting cross-section scales as $\langle \sigma_{1{\rm loop}}v\rangle \sim f
(\alpha_3/4\pi)^2 \lambda^4/s$, where $\lambda$ is the 
spurion-messenger-messenger coupling in the GMSB superpotential, $\alpha_3$
the strong coupling constant and
 $f$ is a prefactor which embodies the model dependence: a very
 small $f$ occurs in the decoupling regime  $ M_S \gg M_X $ 
 ($f \to 0$ when $M_S/M_X \to \infty$),   but a large value
 is also possible close to resonance when $M_X \simeq M_S/2$. However, more
 generally $f \sim {\cal O}(1)$ for $M_S \sim M_X$ away from the resonance.
 As we will stress later on, in the regime where  $\langle \sigma_{1{\rm loop}}v\rangle$
 is very small the lightest messenger freezes-out while relativistic and 
 the relic density becomes independent of the details of the
  cross-section.

In the region of parameter space we are interested in, namely
$m_{3/2}\lesssim 10\,$MeV and $M_X\lesssim 10^7\,$GeV, the
annihilation into goldstinos can be neglected with respect to the
one-loop annihilation cross-section provided $f\gtrsim 6\cdot 10^{-10}
(M_X/10^6\,{\rm GeV})^6(m_{3/2}/1\,{\rm MeV})^{-4}\lambda_{0.1}^{-4}$,
where we have defined $\lambda_{0.1}\equiv (\lambda/0.1)$. 
Note that in this
region of parameter space, the estimate of annihilation cross-section
into a pair of goldstinos remains perturbativily reliable. This is
contrary to most scenarios of gravitino dark matter in $SU(5)$ grand
unification where the solution lies in a region where the
cross-section seems to violate unitarity (i.e. $\sigma v\gtrsim
8\pi/M_X^2$ as $v\to 0$), so that the conclusions depend strongly on
what one assumes for the behavior of this cross-section, see
\cite{JLM04}, and furthermore multi-goldstino production can then
become important to consider.

The relic abundance of the lightest messenger can thus be written as:

\begin{equation}
Y_X \,\simeq\, 8.4\cdot 10^{-10}\,\left({M_X\over 10^6\,{\rm GeV}}\right)
\left({4\pi\over\alpha_3}\right)^2\lambda_{0.1}^{-4}{x_f \over f}\,
\label{eq:YX}
\end{equation}
with $x_f = \log\left[Q_f/\sqrt{\log(Q_f)}\right]$ and $Q_f \simeq
1.5\cdot 10^{6} f (\alpha_3/4\pi)^2 \lambda_{0.1}^4
(M_X/10^6\,{\rm GeV})^{-1}$. The
parameter $x_f$ denotes the ratio of the messenger mass to the temperature
at which freeze-out of annihilations occurs. Hence
\begin{equation}Y_X \simeq
5.0\cdot 10^{-5} f^{-1}\lambda_{0.1}^{-4}
\left({M_X\over10^6\,{\rm GeV}}\right)\left[1 + 0.26\log(f) -
0.26\log\left({M_X\over10^6\,{\rm GeV}}\right)
+1.05\log\left(\lambda_{0.1}\right)\right].
\end{equation}  
When $10^{-2}\lesssim \lambda_{0.1}^{-4}(M_X/10^6\,{\rm GeV})/f\lesssim 1$, 
$Y_X \simeq 5\cdot 10^{-5}
(M_X/10^6{\rm GeV})^{0.8}f^{-0.8}\lambda_{0.1}^{-3.2}$ provides a good approximation. For
other values of $M_X$ and $f$, one needs to keep the logarithmic
correction.

Eventually, $X$ will decay to visible sector particles, otherwise it
would overclose the Universe. As mentioned earlier, we assume that $X$
decay can occur through non-renormalizable operators. The main
motivation for this is to preserve the phenomenological successes of
GMSB models, notably with respect to flavor changing neutral currents
and electroweak symmetry breaking. It is certainly possible to
introduce renormalizable operators that violate messenger number and
yet do not spoil the features of GMSB models, however such operators
have to be suppressed by unnaturally small numerical
prefactors~\cite{DNS97}, particularly for small messenger masses.
Non-renormalizable higher dimension operators are expected to occur at
the Planck scale, which respect the gauge symmetry of $SO(10)$ as well
as $R-$symmetry and possibly other global symmetries. Depending on the
charge assignments [which should forbid dangerous couplings, {\sl
e.g.} between messengers and Higgses with GUT {\sl vev's}], some of
these operators will violate messenger number by one unit. For these
operators, the typical decay width of $X$ is $\Gamma_X \simeq
(1/16\pi) f' M_X^3/m_{\rm Pl}^2$, with $f'$ a numerical factor of
order unity.  Decay occurs when the Hubble rate $H\simeq \Gamma_X$, or
equivalently, at background temperature

\begin{equation}
T^> \simeq 87\,{\rm MeV}\,
f'^{1/2}\left({M_X\over 10^6\,{\rm GeV}}\right)^{3/2}
\left({g_{\star}^>\over 10.75}\right)^{-1/4}.
\label{T>}
\end{equation}

  The decay width of the lightest messenger is so suppressed and its
relic abundance is so large (due to its suppressed annihilation
cross-section) that $X$ actually comes to dominate the total energy
density budget before decaying. This happens since $X$ quanta become
non-relativistic at temperatures $\lesssim M_X$ and therefore their
energy density redshifts less fast than that of radiation. This era of
non-relativistic matter domination starts at background temperature
$T_{\rm dom}\simeq (4/3)M_XY_X\simeq 67\,{\rm GeV}\,
f^{-0.8}\lambda_{0.1}^{-3.2}(M_X/10^6\,{\rm GeV})^{1.8}$, which indeed exceeds $T^>$. The
decay of this lightest messenger thus results in a significant amount
of entropy generation, by reheating the Universe to temperature $T^>$.
The amount of entropy produced can be written in a rather simple way
as~\cite{JLM04}:

\begin{equation} 
\Delta S \,\simeq\, {T_{\rm
dom}\over T^>}\,\simeq\, 7.6\cdot 10^2 f^{-0.8}
\lambda_{0.1}^{-3.2}f'^{-1/2} \left({M_X \over 10^6\,{\rm
GeV}}\right)^{0.3}, \label{eq:DeltaS} \end{equation} and all species are
diluted with respect to the background entropy density by the factor
$\Delta S$.

In particular, the goldstino abundance after lightest messenger decay
reads: 
\begin{equation}
Y_{3/2} \simeq {1\over\Delta S} Y_{3/2}^{\rm eq} \simeq 2.1\cdot10^{-6}
f^{0.8}\lambda_{0.1}^{3.2}f'^{1/2}\left({M_X\over10^6\,{\rm GeV}}\right)^{-0.3}.\label{eq:Y32-1}
\end{equation} 
This gives a present-day relic abundance: 
\begin{equation} \Omega_{3/2}h^2 \,\simeq\, 0.067
f^{0.8} \lambda_{0.1}^{3.2} f'^{1/2}\left({M_X\over 10^6\,{\rm
GeV}}\right)^{-0.3}\left({m_{3/2}\over0.1\,{\rm MeV}}\right).
\label{eq:Omega} \end{equation}
Hence one finds the correct relic abundance for dark matter for
$m_{3/2}\sim 0.1-1\,$MeV, $M_X\sim 10^6\,$GeV and 
$\lambda_{0.1}^4f\sim f'\sim {\cal
O}(1)$. If the one-loop annihilation cross-section prefactor 
$\lambda_{0.1}^4f$ takes
values significantly larger than unity, the region in which one finds
suitable gravitino dark matter shifts down to smaller $m_{3/2}$, and
remains at the same values of $M_X$. Indeed one must recall that the
lower bound $M_X \gtrsim 10^5\,$GeV is imposed by the phenomenology of
GMSB models. Note also that as $M_X$ increases above $\sim 10^7\,$GeV,
depending on $\lambda_{0.1}^4f$ and $m_{3/2}$, annihilation into a pair of goldstinos
may come to dominate the one-loop annihilation channel. Due to the strong 
dependence of the former on $M_X$, 
the relic abundance of $X$ quickly decreases in this case, 
as $M_X$ increases, hence entropy production becomes less
effective and gravitinos tend to overclose the Universe. For 
$\lambda_{0.1}^4f\sim
10^3$, the region in which one finds gravitino dark matter lies at
$m_{3/2}\sim1\,$keV and $M_X \sim 10^6\,$GeV. 

 Conversely, if $\lambda_{0.1}^4f$ is significantly smaller than unity, more
precisely 
$\lambda_{0.1}^4f\lesssim 10^{-1} (M_X/10^6\,{\rm GeV})$, freeze-out of
lightest messenger annihilations can occur as early as in the
relativistic regime, i.e.  $x_f \lesssim 1.5$. In this case, the relic
abundance $Y_X\sim 1.2\times 10^{-3}$ does not depend anymore on the
annihilation cross-section, and the relic gravitino abundance reads:
\begin{equation}
\Omega_{3/2}h^2 \simeq 0.028 f'^{1/2} \left({M_X\over10^6\,{\rm
GeV}}\right)^{1/2}\left({m_{3/2}\over1\,{\rm MeV}}\right).
\end{equation} 
One may still find satisfactory solutions for gravitino dark matter,
if $m_{3/2}\gtrsim 1\,$MeV, $M_X\gtrsim 10^6\,$GeV and/or $f'\gtrsim
1$. Note that $m_{3/2}\lesssim 10\,$MeV is imposed by big-bang
nucleosynthesis constraints on high energy injection if the NLSP is a
bino~\cite{JLM04}; for a stau NLSP, the constraint is generally
weaker, $m_{3/2}\lesssim 5\,$GeV \cite{FST04}.

It is safe to ignore the production of goldstinos during this second
stage of reheating since the corresponding temperature $T^>$ given by
Eq.~(\ref{T>}) is well below the sparticles masses.  Goldstinos are
also produced by the decay of the next-to-lightest supersymmetric
particle (NLSP), which depending on the underlying mass spectrum, may
be generally a bino or a stau. NLSPs result not only during a thermal
freeze-out but possibly also during the messenger decays themselves.
The former contribution to goldstinos is negligible due to the entropy
release that follows NLSP freeze-out; the latter is also generally
small when compared to the previously existing gravitino
abundance. The decay width of the NLSP is $\Gamma_{\rm NLSP}\simeq
(1/48\pi) M_{\rm NLSP}^5/(m_{3/2}^2m_{\rm Pl}^2)$, so that decay
occurs at background temperature $T_{\rm NLSP}\simeq 5\,{\rm
MeV}\,\left({M_{\rm NLSP}/ 100\,{\rm GeV}}\right)^{5/2}
\left({m_{3/2}/ 1\,{\rm MeV}}\right)^{-1}$.  If one assumes that $X$
can produce NLSPs in its decay, the NLSPs produced have time to
annihilate before decaying to goldstinos, depending on the comparison
of $T_{\rm NLSP}$ and $T^>$. To be conservative, one may assume that
$N_{\rm NLSP}$ NLSPs are produced per messenger decay,
with~\cite{FY02}:
\begin{equation}
N_{\rm NLSP}\approx {M_X\over M_{\rm NLSP}^2/T^>}\simeq 5 f'^{1/2}
\left({M_X\over10^6\,{\rm
GeV}}\right)^{5/2}
\left({M_{\rm NLSP}\over100\,{\rm GeV}}\right)^{-2}.
\end{equation} 
If these can decay to goldstinos directly without annihilating, 
the amount of gravitinos produced is:
\begin{equation}
Y_{3/2}^{\rm NLSP}\sim N_{\rm NLSP}{Y_X \over \Delta S} 
\sim 2 \cdot 10^{-7}f'\left({M_X\over 10^6\,{\rm GeV}}\right)^3 \left({M_{\rm
NLSP}\over 100\,{\rm GeV}}\right)^{-2},
\end{equation} 
which remains small compared to $Y_{3/2}$ calculated in
Eq.~(\ref{eq:Y32-1}) above provided $M_X\lesssim 10^6\,$GeV and/or
$M_{\rm NLSP}\gtrsim 100\,$GeV. Moreover NLSP annihilations prior to
NLSP decay, reduce this estimate~\cite{FH02,FY02} to
\begin{equation}
Y_{3/2}^{\rm NLSP}\simeq 1.8\cdot 10^{-11} f'^{-1/2}\left({\langle\sigma_{\rm
NLSP}v\rangle\over 10^{-7}\,{\rm GeV}^{-2}}\right)^{-1}
\left({M_X\over 10^6\,{\rm GeV}}\right)^{-3/2}
\left({g_{\star}^>\over 10.75}\right)^{1/4},
\end{equation}
provided $T_{\rm NLSP} < T^>$, which is the case unless $f'$ is very 
small. Annihilations are quite effective for stau NLSPs
with typical $\langle\sigma_{\rm NLSP}v\rangle\sim 10^{-7}\,{\rm
GeV}^{-2}(m_{\rm NLSP}/100\,{\rm GeV})^{-2}$.

The present scenario respects the various constraints from
big-bang nucleosynthesis (BBN). For instance, the lightest messenger
and the NLSP decay well before nucleosynthesis provided $M_X\gtrsim
10^5\,$GeV and $m_{3/2}\lesssim 10\,$MeV so that their decay products
can thermalize and constraints from high energy injection do not
apply. On the other hand, for gravitino masses $\sim0.1-1\,$GeV decay
occurs during BBN, and for the right NLSP abundance and hadronic
branching ratio, interesting effects on the $^{7}$Li and $^{6}$Li
abundances may occur~\cite{Jeda04}. Nevertheless, such high $m_{3/2}$
are successfully accommodated in the present scenario only for $f'\ll
1$.  The gravitinos produced in NLSP decay, although they remain
highly relativistic at the time of big-bang nucleosynthesis, are in
too small numbers to contribute significantly to the energy
budget. One indeed evaluates the fractional contribution to the
radiation energy density carried by those relativistic gravitinos at
BBN to be $\sim Y_{3/2}^{\rm NLSP}M_{\rm NLSP}/T_{\rm NLSP} \sim
2\cdot10^4\, Y_{3/2}^{\rm NLSP} (M_{\rm NLSP}/100\,{\rm GeV})^{-3/2}
(m_{3/2}/1\,{\rm MeV})$, hence it can be neglected. If $R-$parity does
not hold, the NLSP can decay much earlier through other channels, and
BBN constraints are actually replaced with constraints on the
gravitino lifetime from the non-observation of cosmic diffuse
backgrounds distortions. However, it has been shown that if
$m_{3/2}\lesssim 10\,$MeV, these constraints are
satisfied~\cite{TY00}: the gravitino is then so long-lived that it can
be considered as stable on our cosmological timescale. Hence the
present scenario for gravitino dark matter remains valid even if
$R-$parity is violated. 

Finally, note that the bulk of gravitinos produced here behave as cold
dark matter from the standpoint of large scale structure formation,
even for gravitinos masses as small as $\sim $keV, due to the cooling
of gravitinos during entropy release. One indeed
calculates~\cite{JLM04} that the present-day gravitino velocity
$v_0\sim 0.0017\,{\rm km/s}\,(m_{3/2}/1\,{\rm
keV})^{-1}f^{0.26}\lambda_{0.1}^{1.04}f'^{1/6}(M_X/10^6\,{\rm GeV})^{-0.1}$  for the
dominant component that results from thermal equilibrium (which
excludes the non-thermal part from NLSP decay). The corresponding
smoothing spatial scale for structure formation can be calculated as
$R\sim 235\,{\rm kpc}\, (v_0/0.05{\rm km/s})^{0.86}$~\cite{BHO:01},
which confirms that these gravitinos are approximately cold. This
situation is contrary to that encountered in scenarios in which more
massive dark matter gravitinos $m_{3/2}\sim 10-100\,$GeV are produced
by non-thermal processes (in particular by the decay of the NLSP); in
these models, indeed, the gravitino dark matter is warm in a large
part of parameter space~\cite{K05}.

In conclusion, for a reasonable choice of parameters of the underlying
GMSB model, one can obtain the right amount of gravitinos to explain
the cold dark matter content of the Universe. The vast majority of
these gravitinos have been produced in scatterings (or decays) at high
temperatures and cooled and diluted by the lightest messenger
out-of-equilibrium decay. A significant advantage of the present
scenario is that the final abundance of gravitinos is independent of
the post-inflationary reheating temperature $T_{\rm RH}$. In
particular, the stringent constraints on $T_{\rm RH}$ for light
gravitinos \cite{CHKL99,M95} are irrelevant here. This implies notably
that scenarios of leptogenesis from right-handed (s)neutrino decay can
operate at high temperatures and produce the observed baryon
asymmetry. This possibility has been discussed in some detail in a
related context by Fujii \& Yanagida~\cite{FY02}. Here we omit the
details for simplicity. The decay of each right-handed (s)neutrino
yields a net lepton asymmetry:

$$\vert\epsilon_1\vert \simeq {3\over 8\pi} {M_{R1}m_{\nu3}\over
\langle H_u^0\rangle^2}\delta_{\rm eff},$$ with $M_{R1}$ the lightest
RH (s)neutrino mass, $m_{\nu3}\sim 0.06\,$eV the heaviest left-handed
neutrino mass, $\langle H_u^0\rangle$ a Higgs {\it vev}, and
$\delta_{\rm eff}$ an effective $CP-$violating phase. Assuming that
the RH (s)neutrino is initially in thermal equilibrium (if $T_{\rm
RH}\gtrsim M_{R1}$), the net baryon asymmetry produced
reads~\cite{FY02}:

\begin{equation}
{n_{\rm B}\over s}\,\approx\, 3.6\cdot 10^{-3} {C\over \Delta
S}\vert\epsilon_1\vert\alpha,
\end{equation}
where $C=-8/23$ in the minimal supersymmetric standard model denotes
the effectiveness of $L$ to $B$ conversion, and $\alpha\sim 1$
characterizes the fraction of lepton asymmetry surviving after RH
(s)neutrino decay. One finally obtains:

\begin{equation}
{n_{\rm B}\over s}\,\approx\, 1.4\cdot 10^{-10} f^{0.8} \lambda_{0.1}^{3.2}
f'^{1/2}\left({M_X\over
10^6\,{\rm GeV}}\right)^{-0.3}\left({M_{R1}\over 5. \; 10^{11}\,{\rm
GeV}}\right), \end{equation} for $\delta_{\rm eff}\sim \alpha\sim 1$,
which matches the measured asymmetry $\simeq 8\cdot10^{-11}$.

\section{Conclusions}

We have shown that a gravitino of mass $m_{3/2}\sim 0.1-1\,$MeV
provides a natural cold dark matter candidate in $SO(10)$ GMSB
scenarios coupled to supergravity, with messenger mass scale $M_X\sim
10^6\,$GeV and non-renormalizable messenger-matter coupling. Although
our conclusions are largely independent of the details of the
underlying GMSB model, we find that our scheme can be accommodated in
the so-called indirect ``simplest'' GMSB models {\it \`a la} Dine {\it
et al.}~\cite{GMSB}, in which the dynamical supersymmetry breaking in
a secluded sector is fed down radiatively to the messenger sector. For
a reasonable choice of the various model parameters, e.g. the coupling
constants, one obtains the correct dark matter abundance. Our results,
and in particular the required combination of $m_{3/2}$ and $M_X$,
depend on the numerical prefactor $f$ of the one-loop annihilation
cross-section of the lightest messenger, which is a $SU(3)\times
SU(2)\times U(1)$ singlet scalar, the spurion-messenger coupling constant
$\lambda$,
as well as on $f'$ the prefactor of
the decay width of the lightest messenger via non-renormalizable
messenger-matter interactions.  Non-renormalizable interactions are
preferred over their renormalizable counterparts in order to not
violate existing limits on flavor-changing neutral currents.  On the
other hand, messengers have to decay to be cosmologically acceptable.
When $f\lambda^4$ is sufficiently small, our results do not depend on the
magnitude of the messenger annihilation cross section.

  Our results do not depend on the mass spectrum in the visible
sector. In effect the final relic gravitino abundance is simply the
abundance of goldstinos at thermal equilibrium diluted by the entropy
produced in the lightest messenger late decay. Goldstinos are
initially brought in thermal equilibrium by scatterings and decays
involving sparticles as well as messenger fields. This fact is at
variance with other SUSY models in which the dark matter candidate is
a neutralino LSP; in those models, one must find the right combination
of the parameters that determine the visible sector mass spectrum,
e.g. $m_{1/2}$, $m_0$, ($A_0$), $\tan\beta$ and $\rm sign(\mu)$ to
obtain the right dark matter abundance.

  Gravitino dark matter in this mass range $m_{3/2}\gtrsim\,$keV
cannot be observed in dark matter search experiments, either direct or
indirect. However there exist interesting proposals to detect evidence
for gravitinos/goldstinos~\cite{BHRY04} in next generation
colliders. Such experiments should also confirm or dispute the GMSB
phenomenology which leads to a highly predictive mass spectrum with
distinctive features.

\end{document}